\documentclass[conference]{IEEEtran}

\usepackage{ifpdf}
\usepackage{amssymb}
\usepackage{ifpdf}
\usepackage{float}

\usepackage{cite}
\usepackage{caption}
\usepackage{lipsum}
\ifCLASSINFOpdf
  \usepackage[pdftex]{graphicx}
   \graphicspath{{../pdf/}{../jpeg/}}
  \DeclareGraphicsExtensions{.pdf,.jpeg,.png}
\else
   \usepackage[dvips]{graphicx}
   \graphicspath{{../eps/}}
   \DeclareGraphicsExtensions{.eps}
\fi

\usepackage{algorithmic}

\usepackage{eqparbox}

\usepackage[T1]{fontenc}
\usepackage[utf8]{inputenc}
\usepackage{fancyhdr}
\usepackage{amsmath}
\usepackage{amssymb}
\usepackage{bbold}

\newsavebox{\ieeealgbox}

\begin{document}
%
\title{Model-Driven Deep Learning Method for Jammer Suppression in Massive Connectivity Systems} 

\author{\IEEEauthorblockN{Milutin Pajovic and Toshiaki Koike-Akino and Philip V. Orlik}
\IEEEauthorblockA{Mitsubishi Electric Research Laboratories (MERL) \\ Cambridge, MA, 02139, USA \\ Email: pajovic@merl.com}
}

\maketitle

\begin{abstract}
We present a method for separating collided signals from multiple users in the presence of strong and wideband interference/jamming signal. More specifically, we consider a massive connectivity setup where few, out of a large number of users, equipped with spreading codes, synchronously transmit symbols. The received signal is a noisy mixture of symbols transmitted through users' flat fading channels, impaired by fast frequency hopping jamming signal of relatively large power.
In the absence of any conventional technique suitable for the considered setup, we propose a "model-driven" deep learning method, based on convolution neural network, to suppress jamming signal from the received signal, and detect active users together with their transmitted symbols. A numerical study of the proposed method confirms its effectiveness in scenarios where classical techniques fail. As such, in a two user scenario with wideband jamming signal of power $20$\;dB above the power any active user, the proposed algorithm achieves error rates $10^{-2}$ for a wide range of AWGN variances. 
\end{abstract}

\IEEEpeerreviewmaketitle

\section{Introduction}
\label{sec:intro}

Massive machine type communication (mMTC) has been identified as one of three broad categories of the emerging 5G use cases and scenarios~\cite{5G_overview_JSAC17}. The mMTC embodies a random connectivity setup consisting of hundreds, or even thousands, of users that are in grant-free communication with the same base station over a shared wireless channel and occasionally transmit short messages that collide~\cite{Bockelmann_TETT13}. A variety of methods have been proposed to separate colliding users. As such,~\cite{Shulkind_GlobalSIP17, Pajovic_GC18} consider user messages comprising of the same, known preamble and information-bearing payloads, while~\cite{SWang_CL15, Pajovic_ICC18,Xie_RedDimDet_TIT18} equip users with (non)orthogonal spreading codes to facilitate their separation. Commonly, the proposed methods leverage sparsity in the user activity domain to separate colliding users. 

Separation of colliding users in the presence of an interference originating from other nearby systems or malicious jammers has received much less attention despite being of high practical importance. For example, a communication system with a large number of IoT devices may be maliciously jammed to the extent that messages delivering critical sensor measurements cannot be recovered. Similar to recent work~\cite{Jung_AntiJamming_TAES18}, we consider a massive connectivity scenario with synchronous users equipped with spreading codes whose transmissions are subject to a relatively strong wideband jammer employing fast frequency hopping. In comparison to~\cite{Jung_AntiJamming_TAES18}, where users employ different spreading codes over symbol periods, we assume each user is assigned one spreading code and suppress the jamming signal using a deep learning (DL)-based method. Viewing the jammer suppression problem as the one similar in nature to image denoising, we propose a DL architecture inspired by state-of-the-art image denoising method~\cite{GaussImageDenois_Zhang_TIP17} and further enhance it by incorporating domain-specific knowledge in its design. Finally, assuming the DL method yields a jammer-free signal, we employ reduced dimension decorrelating (RDD) algorithm~\cite{Xie_RedDimDet_TIT18} to separate active users. The proposed method is tested with simulations which reveal that error rates around $1$\% can be achieved in scenarios of two colliding users of equal power, jammed with a wideband signal 20\;dB above the level of any of the active users.

\section{RELATION TO PRIOR WORK}
\label{sec:prior}

A growing research body on grant-free massive connectivity systems consider different scenarios and employ a variety of sparsity-based recovery methods for user separation. As such, element-wise~\cite{Bockelmann_TETT13}, group-wise~\cite{Xie_GropWise_TComms16} and analog sparsity~\cite{Xie_RedDimDet_TIT18, Pajovic_ICC18} are used to separate time and frequency synchronized users whose channels are known on the receiver side. Approximate message passing has been leveraged to estimate active users and their channels~\cite{Sun_MassiveConnectivity_GC18}, as well as the transmitted symbols~\cite{Ding_MUD_mMTC_Jul18}. A more challenging time and/or frequency asynchronous systems are considered in~\cite{Li_LinkAcq_TIT13}, where user-specific messages are transmitted for link acquisition over unknown multi-path channels;~\cite{Chi_SSP12}, where users are equipped with spreading codes and channel state information is available on the receiver side;~\cite{SWang_CL15}, where users also employ spreading codes, their multi-path channels are unknown and maximum delay spread is one symbol duration; and~\cite{Pajovic_GC18}, where packets transmitted over unknown channels are comprised of the same, known preamble and random payload, and experience phase noise impairments. Nevertheless, none of these works consider a fairly realistic scenario where user transmissions are subject to interference from other systems or malicious jammers.

Considering the setup, the closest works to this paper are~\cite{Raju_ICA_TWC06} and~\cite{Jung_AntiJamming_TAES18}. As such,~\cite{Raju_ICA_TWC06} uses independent component analysis (ICA) to separate users equipped with spreading codes of unknown channels. However, one practical limitation of that approach is that the maximum number of users $N$ that a single antenna ICA receiver can separate is $N<2S/3-1$, where $S$ is the spreading factor. A combination of the robust principal component analysis (PCA) and ICA to separate users employing spreading codes in the presence of a jamming signal is proposed in~\cite{Jung_AntiJamming_TAES18}. This system assumes that each user applies different spreading codes over symbol periods within the transmitted packet, so that the received signal is represented in a form that facilitates jammer suppression and user separation using the robust PCA. 
Similar to~\cite{Raju_ICA_TWC06} and~\cite{Jung_AntiJamming_TAES18}, we also consider a fast frequency hopping jammer that imperils transmissions from multiple users. In comparison to those works, each user in our setup is assigned a fixed spreading code that does not change over symbol periods and the number of users is not bounded by some fraction of the spreading factor $S$, thereby leaving the robust PCA and ICA techniques inapplicable for jammer suppression and user separation.

Since user separation in the presence of jamming signal in our setup is not amenable to conventional signal processing methods, we resort to DL framework with the goal to suppress the jammer and enable detection of active users and their symbols. Despite its enormous success in other fields, DL has recently started to gain interest as a supplement or alternative to other approaches in communications~\cite{Wang_DNNforComms_ChinaComms17,OShea_TCCN17}. To the best of our knowledge, DL has not been applied for jammer suppression or, more generally, in massive connectivity systems where users employ spreading codes. The DL architecture used in the proposed method is a variation of the one used in state-of-the-art DL method for image denoising~\cite{GaussImageDenois_Zhang_TIP17}, and further enhanced by including domain-specific knowledge into its design. The resulting algorithm can be viewed to as an instance of a framework recently cast as model-driven DL~\cite{He_ModelDrivenDL_Sept18}.

\section{Signal Model}
\label{sec:model}

 

In the considered setup, $N$ users are in communication with a common base station/access point over a shared wireless channel. The users are equipped with spreading codes of spreading factor $S$. As such, user $i$'s chip-rate discrete-time domain representation of the spreading code is ${\bf s}_i \in \mathbb{C}^{S \times 1}$. The spreading codes are arranged into a matrix of spreading codes, $\mathbf{S} \in \mathbb{C}^{S \times N}$, such that its $i$-th column is ${\bf s}_i$.
We assume the users are in perfect synchronization, meaning that they have a common time reference as to when a symbol time starts and ends. Each user sends a (possibly) precoded data symbol $x_i$ which undergoes a frequency flat fading channel $h_i \in \mathbb{C}$ before it reaches the receiver. 
Since users share communication channel, collisions occur when two or more users send information during the same symbol period. The mixture of the signals transmitted during one symbol period is then given by
\begin{equation}
{\bf y} = \mathbf{H} \mathbf{S} \mathbf{x} \label{mixed_sig}
\end{equation} 
where $\mathbf{H}$ is a diagonal matrix of users' channels such that $\left[ \mathbf{H} \right]_{ii}=h_i$, while ${\bf x} \in \mathbb{C}^{N \times 1}$ is a vector of transmitted symbols from all users such that if a user $j$ is inactive, its corresponding transmitted symbol $x_j=0$. 

In addition to perfect time synchronization, another common approach in grant-free massive connectivity system is to assume that all user channels are known at the receiver side. We relax this assumption using the approach from~\cite{Pajovic_ICC18}. In short, channel reciprocity inherent to the time domain duplex (TDD) system implies that the channel between the BS and user $i$ is equal to the channel in the opposite direction, $h_i$. Therefore, the BS broadcasts known pilot symbols that user $i$ receives and exploits to estimate its channel $h_i$. Due to time synchronization, all users are able to estimate their channels in parallel. Assuming perfect channel estimation, each user $i$ precodes its symbol $b_i$ with normalized zero-forcing (ZF) precoder such that its transmitted data $x_i$ before spreading is given by
\begin{equation}
x_i = b_i p_i = b_i \frac{h_i^*}{|h_i|} \label{precoded_sym}
\end{equation}
where $p_i$ is the precoder. Substituting~(\ref{precoded_sym}) into~(\ref{mixed_sig}) yields
\begin{equation}
{\bf y} = \tilde{\mathbf{H}} \mathbf{S} \mathbf{b} \label{precoded_mixed_sig}
\end{equation}
where $\tilde{\mathbf{H}}$ is a diagonal matrix of user channels' magnitudes, i.e., $\tilde{\mathbf{H}}=\text{diag}\left(|h_1|,|h_2|, \ldots, |h_N| \right)$, while ${\bf b}$ is a vector of transmitted symbols. An inactive user is assumed to transmit zero symbol and thus $b_i \in \mathcal{M} \cup \{0\}$, where $\mathcal{M}$ is the symbol alphabet for the utilized modulation format. 

The transmitted signals from active users are impaired by a wideband jamming signal. Similar to~\cite{Jung_AntiJamming_TAES18}, the jamming signal is modeled as a fast frequency hopping signal, meaning that the jammer quickly jumps over different and randomly chosen frequencies during one symbol period. Formally, the chip rate samples of the jamming signal are stacked into a vector ${\bf z} \in \mathbb{C}^{S \times 1}$, given by
\begin{equation}
{\bf z} = \left[ A e^{j 2 \pi f_k (k-1) + \phi_k} \right]_{k=1}^S \label{jam_sig}
\end{equation}
where $A$ is the magnitude, and $f_k \sim \text{U}[0,1]$, $\phi_k \sim \text{U}[0,2\pi)$ are, respectively, uniformly sampled jammer's frequency and phase at the time instant of the $k$-th chip. In the most extreme scenario, $f_k$ and $\phi_k$ change with each $k$. Alternatively, $f_k$ and $\phi_k$ are piece-wise constant over a certain number of chips, which is the case in our numerical study. 

Accounting for noise, the chip-rate sampled received signal is finally given by
\begin{equation}
{\bf r} = {\bf y} + {\bf z} + {\bf v} \label{rx_sig}
\end{equation}
where ${\bf v}$ is circularly symmetric additive white Gaussian noise of variance $\sigma^2$, i.e., ${\bf v} \sim \mathcal{CN}(0,\sigma^2 \mathbf{I}_S)$.

\section{Proposed Algorithm}
\label{sec:method}

We describe in this part the proposed algorithm. In short, the received signal is processed through a cascade of two blocks. The aim of the first block is to suppress jamming signal and output a clean mixture of signals transmitted by active users. This mixture is then an input to the second block which outputs the indices of active users, along with their detected symbols. 

\subsection{DL-based Jammer Suppression}

\begin{figure}[htb]
\begin{minipage}[b]{1.0\linewidth}
  \centering
  \centerline{\includegraphics[scale=0.44]{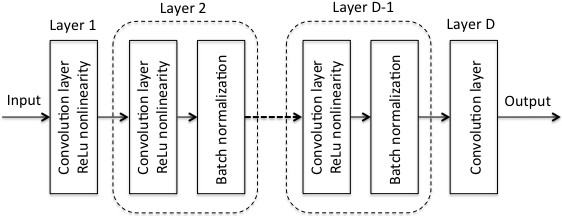}}
\end{minipage}
\caption{DL architecture for jammer suppression.}
\label{fig:DL model}
\end{figure}

The jammer suppression method is designed by learning a discriminative model that maps a pre-processed received signal ${\bf r}$ to the desired mixture of active users' signals ${\bf y}$. To guide the selection of a DL model, we view jammer suppression problem as analogous in nature to image restoration problem. The image restoration problem has received a considerable attention in the literature, and consequently we reduce the search space for DL models by focusing on models that have proven successful for image restoration, and, in particular, image denoising tasks. Building upon the state-of-the-art image denoising model~\cite{GaussImageDenois_Zhang_TIP17}, our jammer suppression block implements DL model shown in Fig.~\ref{fig:DL model}. The model consists of $D$ layers whose building blocks are convolution sublayer, 
rectified linear unit (ReLu) activation, 
and batch normalization. 
More specifically, the first layer processes the input signal through convolution sublayer, followed by ReLu activation. Each subsequent layer $d$, $d=2, \ldots, D-1$, employs BN to the ReLu activated output from the corresponding convolution sublayer. Finally, a single convolution sublayer is the output layer in the proposed DL model.

\begin{figure}[htb]
\begin{minipage}[b]{1.0\linewidth}
  \centering
  \centerline{\includegraphics[scale=0.3]{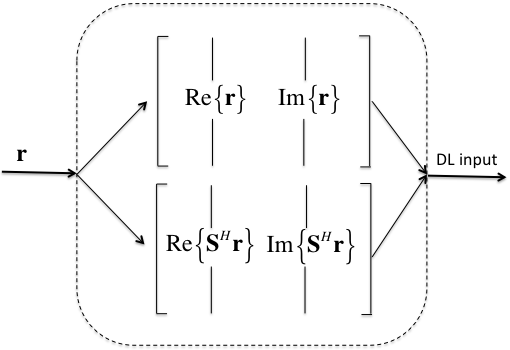}}
\end{minipage}
\caption{Input tensor to the DL model.}
\label{fig:DL input}
\end{figure}

The input to the proposed DL architecture is a tensor consisting of two channels, as depicted in Fig.~\ref{fig:DL input}. The first channel $\tilde{\bf r}$ is formatted into a two-column matrix which contains real and imaginary parts of the received signal $\bf r$. The second channel $\tilde{\bf r}^{\prime}$ is also formatted into a two-column matrix, with real and imaginary parts of the match filter bank (MFB) output ${\bf r}_{\text{MFB}}$ in its columns. In particular, the received signal ${\bf r}$ is processed through the MFB consisting of $S$ filters, each tuned to one spreading sequence ${\bf s}_j$, $j=1,\ldots,S$,
\begin{equation}
{\bf r}_{\text{MFB}}=\mathbf{S}^H {\bf r}
\end{equation}
For an intuitive justification behind the incorporation of the MFB output in the DL input tensor, we recall that the first convolution layer computes weighted combination of the neighboring entries in the input tensor over a patch whose dimension is given by the size of the convolution kernel. In particular, the entries from the $i$th row of the first channel correspond to the $i$th chip's samples of the received signal, while the $i$th row of the second channel is the noisy sum of the user $i$'s transmitted symbol and weighted combination of the jamming signal's chip-rate samples, where the weights are the entries in the spreading code ${\bf s}_i$. Thus, the first convolution layer combines several consecutive samples of the jamming signal and the weighted combinations of the entire jamming signal, as well as the same number of consecutive samples of the mixture of active users' signals and the symbols transmitted by active users. The remaining convolution layers compute the weighted combinations of entries from their input tensors, such that the whole model combines "local" information, embodying consecutive samples of the useful signal ${\bf y}$ and jamming signal ${\bf z}$, with the "global" information, comprising of the transmitted symbols $x_j$ and jamming signal projected onto spreading codes, ${\bf s}_j^H {\bf z}$, with the goal to suppress the jamming signal ${\bf z}$ and yield the mixture of active users' signals ${\bf y}$. We emphasize that the proposed DL model belongs to the class of model-driven DL, whereby enhancing a black-box DL model with communication-specific domain knowledge results in improved performance~\cite{He_ModelDrivenDL_Sept18}.

As a side remark, we note that since the received signal is a jammed version of the useful signal, the input tensor from Fig.~\ref{fig:DL input} resembles a distorted two-channel image and, hence, the expectation is that the DL architecture yielding state-of-the-art image denoising performance is also a promising approach for de-jamming the received signal.

Having specified the DL architecture, its weights ${\bf w}$ are estimated by minimizing the squared $l_2$ norm of the error,
\begin{equation}
\hat{\bf w} = \arg\min_{{\bf w}} \sum_{k=1}^K \| {\bf y}_k - \mathcal{R}({\bf r}_k; {\bf w})\|^2 \label{cost_fun}
\end{equation}
where $K$ is the number of training examples $({\bf r}_k, {\bf y}_k)$, generated according to~(\ref{precoded_mixed_sig}) and~(\ref{jam_sig}), while $\mathcal{R}(\cdot;{\bf w})$ denotes the mapping from the received signal ${\bf r}$ to the DL output, parameterized by the weights ${\bf w}$. 


\subsection{Symbol Detection}

The recovery of active users and transmitted symbols is described in this part. The output from the jammer suppression block is first mapped into a complex domain signal $\tilde{\bf y} \in \mathbb{C}^{S \times 1}$. Assuming the jammer suppression does not distort the mixture of the transmitted users' signals, the output $\tilde{\bf y}$ is given by
\begin{equation}
\tilde{\bf y} = \tilde{\mathbf{H}} \mathbf{S} {\bf b} + \tilde{\bf v} \label{output_model}
\end{equation}
where $\tilde{\bf v}$ models the residual (unsuppressed) jamming signal and noise. The problem of detecting active users and their transmitted symbols from~(\ref{output_model}) is essentially a sparse recovery problem because only few users transmit at the same time, resulting in a sparse vector of their symbols $\bf b$.

We employ here the reduced dimension decorrelating algorithm (RDD)~\cite{Pajovic_ICC18,Xie_RedDimDet_TIT18} to detect $\bf b$ from~(\ref{output_model}). The RDD is a simple, one-shot symbol detection method which cross-correlates $\tilde{\bf y}$ with each spreading code ${\bf s}_i$ to generate statistics $t_i$ corresponding to the $i$th user. Assuming the number of active users $K$ is known (i.e., obtained by some other algorithm), the RDD detects active users from the indices of $K$ largest magnitudes $\{|t_i|\}_{i=1}^N$. Finally, the RDD detects the transmitted symbol of an active user by finding the constellation symbol closest in the Euclidean sense to $t_i$. We note that when spreading codes are orthogonal, the RDD becomes a conventional match filter bank (MFB). 

As a final remark, our choice of the RDD algorithm is motivated by its simplicity and the fact that the focus of this work is suppression of jamming signal. Consequently, the RDD does not limit the generality of the proposed methodology and other sparse recovery algorithms can be used to detect $\bf b$ from $\tilde{\bf y}$.

\section{Simulation Study}
\label{sec:sims}

We consider a simulation scenario where each of $N=128$ users is equipped with orthogonal spreading codes so that the spreading factor $S=128$. The users employ quadrature phase shift keying (QPSK) modulation and precode the symbols using normalized ZF precoder. The signal transmitted from an active user has unit power and experiences a channel with magnitude $|h_i| \sim \text{U}[0.5, 1.5]$. The normalized frequency and phase of the simulated jamming signals, uniformly at random sampled from $[0,1]$ and $[0,2\pi)$, respectively, exhibit 100 jumps over 128 chips, implying that each frequency-phase combination in the jamming signal lasts on average 1-2 chips.

The DL architecture from Fig.~\ref{fig:DL model} implements $D=5$ layers. Each convolution sublayer comprises of 32 convolution filters (i.e., yields 32 feature maps), with the exception of the last layer which implements a single convolution filter. The convolution kernel in each layer covers 5 consecutive rows and both columns (i.e., has dimension 5-by-2) and zero padding is used. Each channel in the input tensor is normalized with the maximum power of the corresponding signal. The RDD processor boils down to a conventional MFB due to orthogonality of the spreading codes. Assuming the number of active users is known, the MFB outputs are used to detect indices of active users and their transmitted QPSK symbols.

We trained and tested the architecture from Fig.~\ref{fig:DL model} with more layers and/or 16, 64 or 128 filters per convolution sublayer, without observing a considerably improved performance. Also, the tests with models where the BN is employed before ReLu show performance deterioration by $\sim$0.5\%, compared to the case when the BN acts upon the ReLu activations, as is the case in the used DL model. We note that our 5-layer model is not as deep as those models where the BN was initially suggested for as a regularizer. However, the tests without BN yield slightly degraded performance, which is the reason why we keep BN in our model. 

The  jammer suppression method is implemented using TensorFlow framework~\cite{TensorFlow}. The DL model parameters, corresponding to a given jamming signal magnitude $A$ and AWGN variance $\sigma^2$, are learned using the ADAM algorithm with batch size 64 and 200 epochs over synthetically generated 200,000 training data points. The learned model is tested with 10,000 Monte-Carlo runs and error rate, defined as a ratio of runs with wrong estimate of an active user index or transmitted symbol, is computed. In other words, a successful run is declared only when all active user indices and their transmitted symbols are correctly detected. The conventional MFB is used as a benchmark since, to the best of our knowledge, we are not aware of any other jammer suppression method suited for the considered setup. 

\begin{figure}[htb]
\begin{minipage}[b]{1.0\linewidth}
  \centering
  \centerline{\includegraphics[scale=0.11]{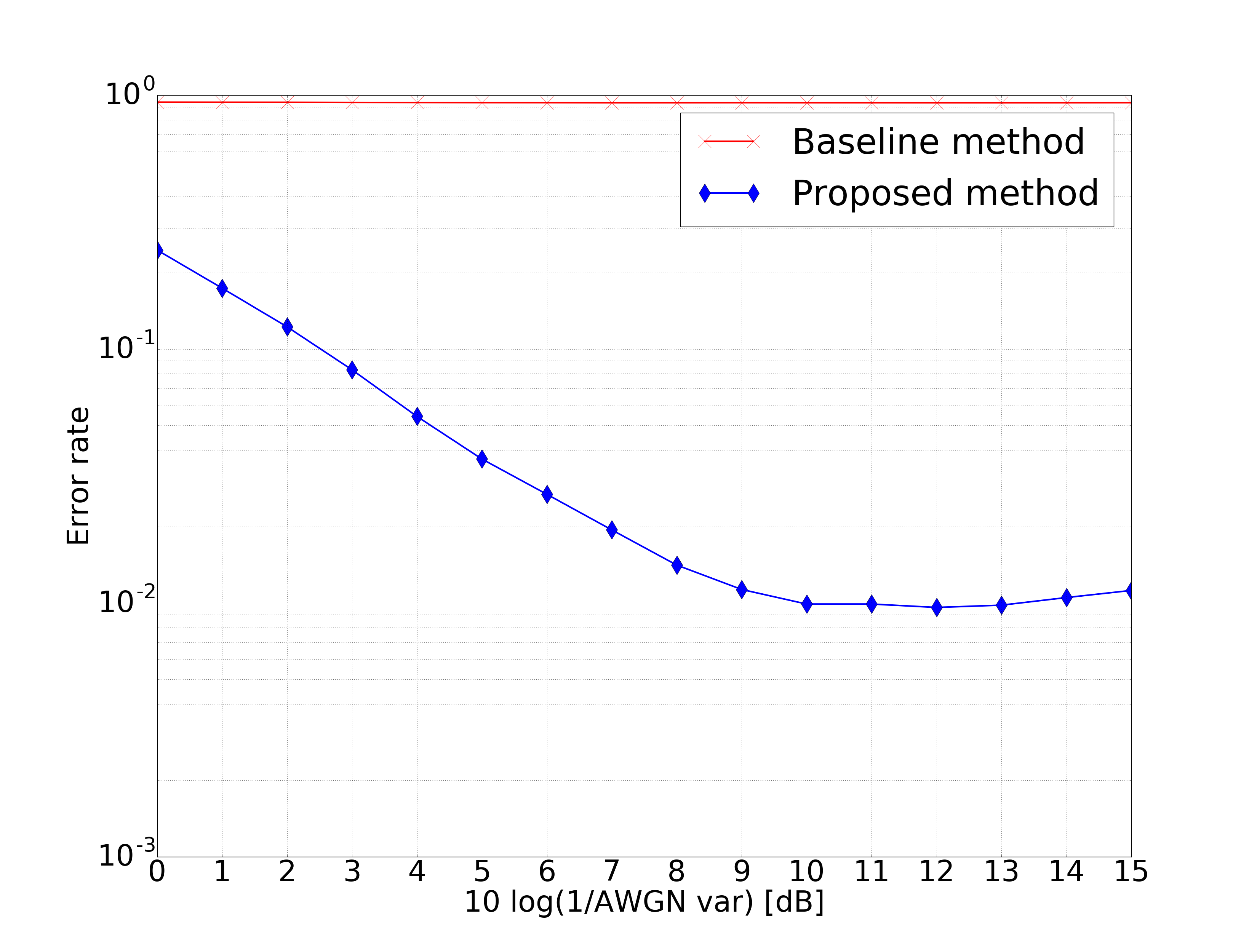}}
\end{minipage}
\caption{Error rate versus AWGN variance for a model trained for jammer power $20$\;dB, AWGN variance $-10$\;dB and two active users.}
\label{fig:AWGN var}
\end{figure}

\begin{figure}[htb]
\begin{minipage}[b]{1.0\linewidth}
  \centering
  \centerline{\includegraphics[scale=0.11]{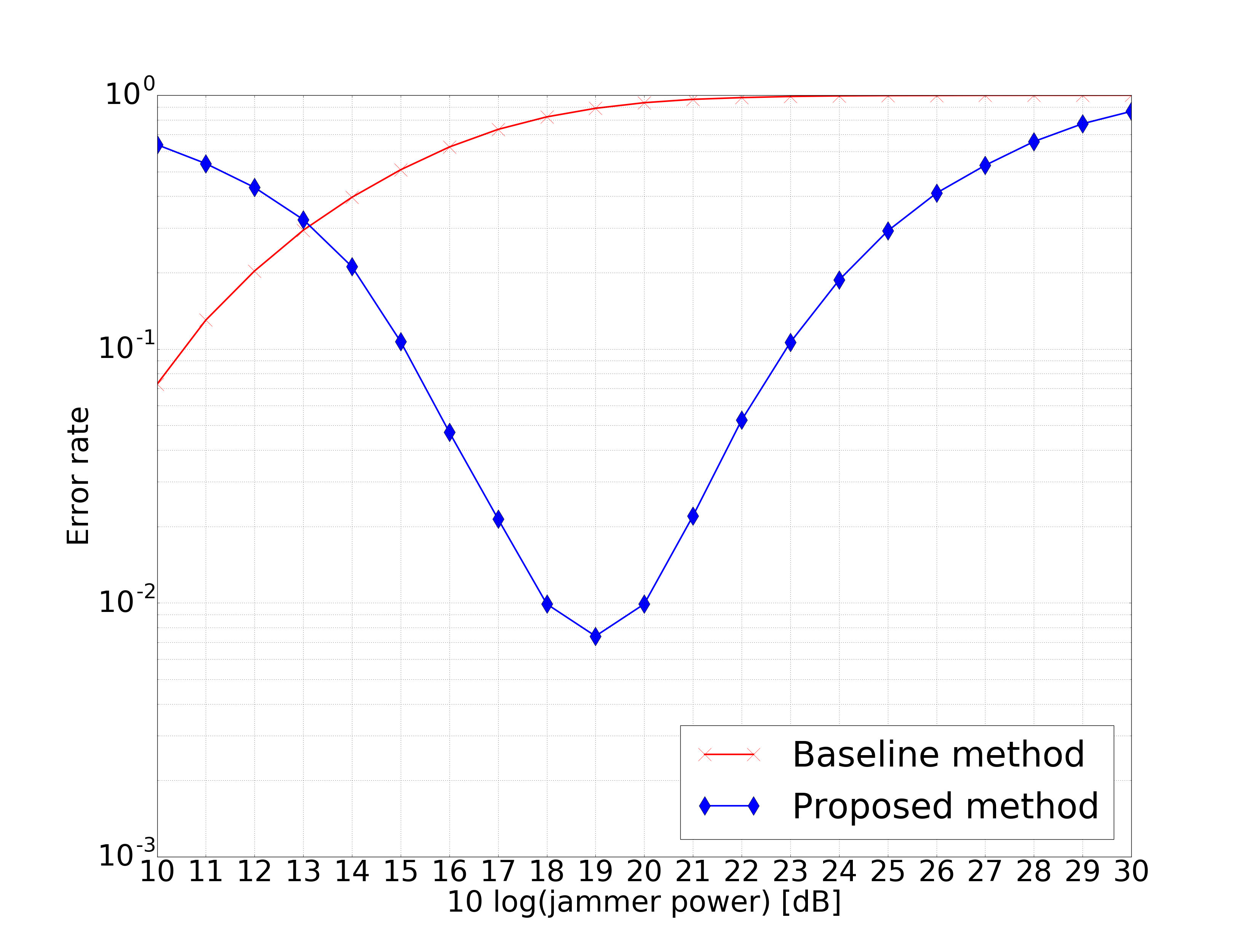}}
\end{minipage}
\caption{Error rate versus jammer power for a model trained for jammer power $20$\;dB, AWGN variance $-10$\;dB and two active users.}
\label{fig:jamming power}
\end{figure}

The error rate dependence on AWGN variance for a model trained for jamming signal power $20$\;dB and noise variance $-10$\;dB with respect to an active user's transmitted signal is shown in Fig.~\ref{fig:AWGN var}. As can be seen, in comparison to the conventional MFB which yields error rates $\sim 95$\%, the error rates of the proposed algorithm are around $1$\% for a smaller and more practical AWGN variances. In the next test, the AWGN variance is fixed to $-10$\;dB and jamming signal power is varied. The resulting error rate performance, shown in Fig.~\ref{fig:jamming power}, confirms that the proposed algorithm considerably outperforms the MFB for stronger jammers.

\begin{figure}[htb]
\begin{minipage}[b]{1.0\linewidth}
  \centering
  \centerline{\includegraphics[scale=0.11]{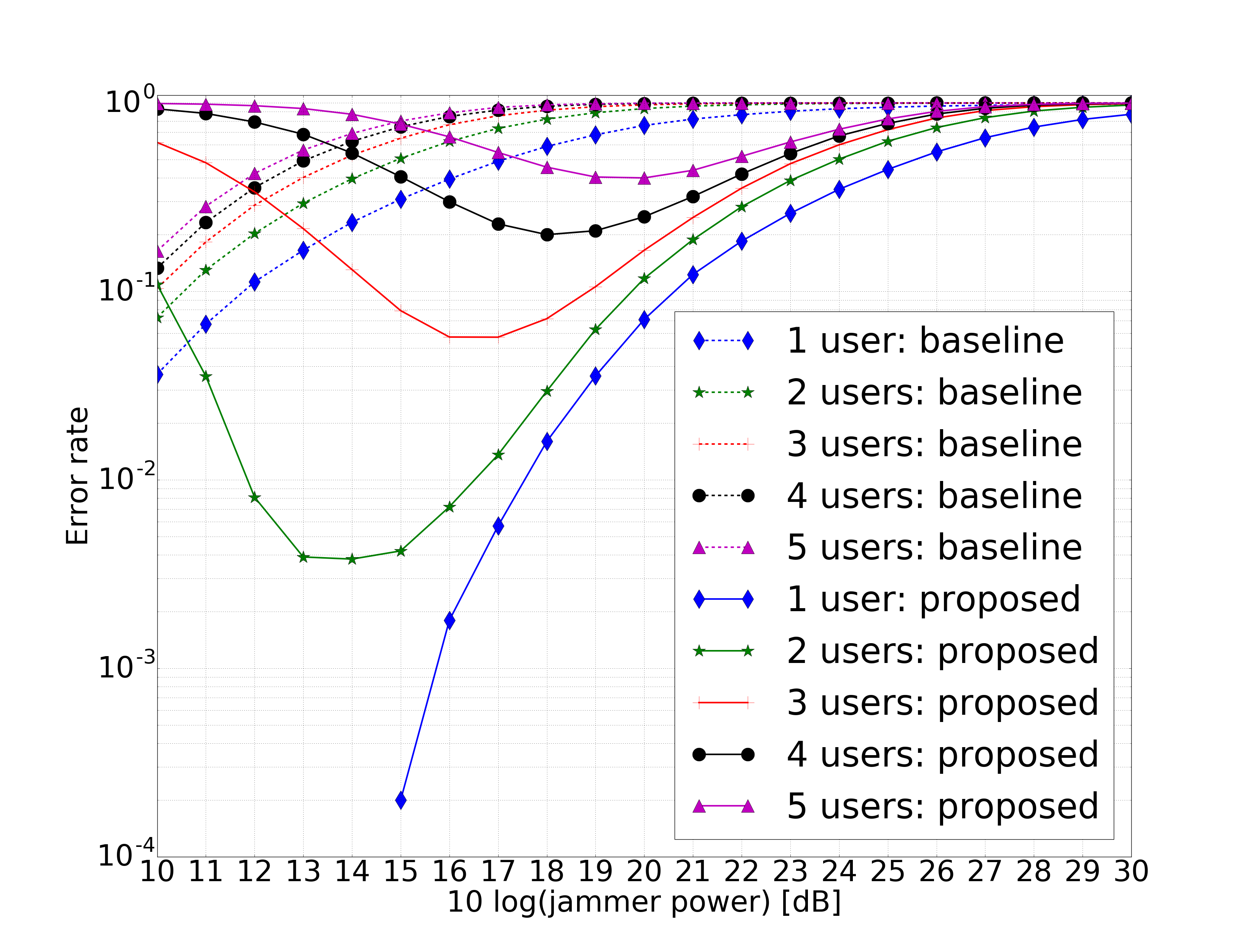}}
\end{minipage}
\caption{Error rate versus jammer power for a model trained for jammer power $20$\;dB, AWGN variance $-10$\;dB and five active users.}
\label{fig:diff number users}
\end{figure}

The error rate performance of the proposed algorithm and MFB benchmark for jammer power sweep and different number of active users is shown in Fig.~\ref{fig:diff number users}. The underlying DL model is trained for 5 active users, jammer power $20$\;dB and AWGN variance $-10$\;dB. We note that the test done in a scenario of one active user does not yield any error over 10,000 runs for jammer powers between $10$ and $14$\;dB. As expected, the error rate performance deteriorates with the increasing number of users. However, even though the model is trained for 5 active users, the error rates are below $10$\% in scenarios with up to three active users and for most considered jammer powers.

\section{Conclusion} \label{Sec:Conclusion}
We consider in this work a massive connectivity system where users are equipped with spreading codes and their transmissions are subject to interference from a relatively strong wideband jamming signal. The jamming signal is suppressed using a model-driven deep learning method that is based on convolution neural network and takes as inputs the received signal and the output from the match filter bank. The active users and their symbols are detected from the de-jammed signal using sparse recovery. The simulation study validates the effectiveness of the proposed method in scenario where conventional techniques fail. As such, error rates of around $1$\% are achieved in scenarios with 2 active users whose signals are impaired by wideband jammer of power $20$\;dB above their individual power levels.




%

\bibliographystyle{IEEEbib}
\bibliography{strings}

\begin{thebibliography}{10}

\bibitem{5G_overview_JSAC17}
M.~Shafi, A.~F. Molisch, P.~J. Smith, T.~Haustein, P.~Zhu, P.~D. Silva,
  F.~Tufvesson, A.~Benjebbour, and G.~Wunder,
\newblock ``5{G}: {A} tutorial overview of standards, trials, challenges,
  deployment, and practice,''
\newblock {\em IEEE Journal on Selected Areas in Communications}, vol. 35, no.
  6, pp. 1201--1221, June 2017.

\bibitem{Bockelmann_TETT13}
C.~Bockelmann, H.~F. Schepker, and A.~Dekorsy,
\newblock ``Compressive sensing based multi-user detection for
  machine-to-machine communication,''
\newblock {\em Transactions on Emerging Telecommunications Technologies}, vol.
  24, no. 4, pp. 389--400, 2013.

\bibitem{Shulkind_GlobalSIP17}
G.~Shulkind, M.~Pajovic, and P.~V. Orlik,
\newblock ``Packet separation in random access channels via approximate sparse
  approximation,''
\newblock {\em IEEE Global Conference on Signal and Information Processing
  (GlobalSIP)}, November 2017.

\bibitem{Pajovic_GC18}
M.~Pajovic, G.~Ozcan, T.~Koike-Akino, P.~Wang, and P.~V. Orlik,
\newblock ``Packet separation in phase noise impaired random access channel,''
\newblock {\em IEEE Global Conference on Communications (Globecom)}, December
  2018.

\bibitem{SWang_CL15}
S.~Wang, S.~An, X.~Miao, Y.~Ma, and S.~Luo,
\newblock ``Compressed sensing assisted joint channel estimation and detection
  for {DSCDMA} uplink,''
\newblock {\em IEEE Communication Letters}, vol. 19, no. 10, October 2015.

\bibitem{Pajovic_ICC18}
M.~Pajovic and P.~V. Orlik,
\newblock ``Reduced-dimension symbol detection in random access channel,''
\newblock {\em IEEE International Conference on Communications (ICC)}, May
  2018.

\bibitem{Xie_RedDimDet_TIT18}
Y.~Xie, Y.~C. Eldar, and A.~Goldsmith,
\newblock ``Reduced-dimension multiuser detection,''
\newblock {\em IEEE Transactions on Information Theory}, vol. 59, no. 6, pp.
  3858?3874, 2013.

\bibitem{Jung_AntiJamming_TAES18}
H.~Jung, K.~Kim, J.~Kang, T.~S. Lee, and S.~Kim,
\newblock ``An i{ALM-ICA}-based anti-jamming {DS-CDMA} receiver for {LMS}
  systems,''
\newblock {\em IEEE Transactions on Aerospace and Electronic Systems}, , no. 5,
  pp. 2318--2328, October 2018.

\bibitem{GaussImageDenois_Zhang_TIP17}
K.~Zhang, W.~Zuo, Y.~Chen, D.~Meng, and L.~Zhang,
\newblock ``Beyond a gaussian denoiser: Residual learning of deep cnn for image
  denoising,''
\newblock {\em IEEE Transactions on Image Processing}, vol. 26, no. 7, pp.
  3142--3155, July 2017.

\bibitem{Xie_GropWise_TComms16}
R.~Xie, H.~Yin, X.~Chen, and Z.~Wang,
\newblock ``Many access for small packets based on precoding and sparsity-aware
  recovery,''
\newblock {\em IEEE Transactions on Communications}, vol. 64, no. 11, pp.
  4680--4694, 2016.

\bibitem{Sun_MassiveConnectivity_GC18}
Z.~Sun, Z.~Wei, L.~Yang, J.~Yuan, X.~Cheng, and L.~Wan,
\newblock ``Exploiting transmission control for joint user identification and
  channel estimation in massive connectivity,''
\newblock {\em IEEE Global Conference on Communications (Globecom)}, December
  2018.

\bibitem{Ding_MUD_mMTC_Jul18}
T.~Ding, X.~Yuan, and S.~C. Liew,
\newblock ``Sparsity learning based multiuser detection in grant-free
  massive-device multiple access,''
\newblock {\em arXiv}, Jul 2018.

\bibitem{Li_LinkAcq_TIT13}
X.~Li, A.~Rueetschi, A.~Scaglione, and Y.~C. Eldar,
\newblock ``Compressive link acquisition in multiuser communications,''
\newblock {\em IEEE Transactions on Information Theory}, vol. 61, no. 12, June
  2013.

\bibitem{Chi_SSP12}
Y.~Chi, Y.~Xie, and R.~Calderbank,
\newblock ``Compressive demodulation of mutually interfering signals,''
\newblock {\em IEEE Statistical Signal Processing Workshop}, 2012.

\bibitem{Raju_ICA_TWC06}
K.~Raju, T.~Ristaniemi, J.~Karhunen, and E.~Oja,
\newblock ``Jammer suppression in {DS-CDMA} arrays using independent component
  analysis,''
\newblock {\em IEEE Transactions on Wireless Communications}, vol. 5, no. 1,
  January 2006.

\bibitem{Wang_DNNforComms_ChinaComms17}
T.~Wang, C.~K. Wen, H.~Wang, F.~Gao, T.~Jiang, and S.~Jin,
\newblock ``Deep learning for wireless physical layer: Opportunities and
  challenges,''
\newblock {\em China Communications}, pp. 92--111, 2017.

\bibitem{OShea_TCCN17}
T.~O'Shea and J.~Hoydis,
\newblock ``An introduction to deep learning for the physical layer,''
\newblock {\em IEEE Transactions on Cognitive Communications and Networking},
  vol. 3, no. 4, pp. 563--575, Dec 2017.

\bibitem{He_ModelDrivenDL_Sept18}
H.~He, S.~Jin, C.~K. Wen, F.~Gao, G.~Y. Li, and Z.~Xu,
\newblock ``Model-driven deep learning for physical layer communications,''
\newblock {\em arXiv}, September 2018.

\bibitem{TensorFlow}
M.~Abadi et~al,
\newblock ``Tensorflow: A system for large-scale machine learning,''
\newblock in {\em 12th USENIX Symposium on Operating Systems Design and
  Implementation (OSDI 16)}, 2016, pp. 265--283.

\end{thebibliography}

\end{document}